\documentclass[vecphys]{svmult}

\usepackage{mathptmx}       
\usepackage{helvet}         
\usepackage{courier}        
\usepackage{type1cm}        
\usepackage{makeidx}         
\usepackage{graphicx}        
\usepackage{multicol}        
\usepackage[bottom]{footmisc}

\makeindex             

\usepackage{color} 

\newcommand{\OmegaM}{\Omega_{{\rm M}}}

\newcommand{\omegaopt}{\omega_{{\rm opt}}}

\newcommand{\mass}{m_{{\rm eff}}}

\newcommand{\xZPF}{x_{{\rm ZPF}}}

\newcommand{\bh}{\hat{b}}

\newcommand{\ah}{\hat{a}}

\newcommand{\nth}{\bar{n}_{{\rm th}}}

\newcommand{\ncav}{\bar{n}_{{\rm cav}}}

\newcommand{\gom}{g}

\newcommand{\GammaM}{\Gamma_{{\rm M}}}

\newcommand{\GammaOpt}{\Gamma_{{\rm opt}}}

\newcommand{\GammaEff}{\Gamma_{\rm eff}}

\newcommand{\cavlength}{L}

\newcommand{\finesse}{\mathcal{F}}

\newcommand{\Pin}{P_{{\rm in}}}

\newcommand{\Langevin}{\xi}

\begin{document}

\title*{Hybrid Mechanical Systems}

\author{Philipp Treutlein, Claudiu Genes, Klemens Hammerer, Martino Poggio, \\and Peter Rabl}

\institute{Philipp Treutlein \at University of Basel, Department of Physics, \email{philipp.treutlein@unibas.ch} \and 
Claudiu Genes \at University of Innsbruck, Institute for Theoretical Physics, \email{Claudiu.Genes@uibk.ac.at} \and
Klemens Hammerer \at University of Hannover, Institute for Theoretical Physics and Institute for Gravitational Physics, \email{Klemens.Hammerer@itp.uni-hannover.de} \and
Martino Poggio \at University of Basel, Department of Physics, \email{martino.poggio@unibas.ch} \and
Peter Rabl \at Vienna University of Technology, Institute of Atomic and Subatomic Physics, \\ \email{peter.rabl@ati.ac.at}
}

\maketitle

\abstract{We discuss hybrid systems in which a mechanical oscillator is coupled to another (microscopic) quantum system, such as trapped atoms or ions, solid-state spin qubits, or superconducting devices. 
We summarize and compare different  coupling schemes and describe first experimental implementations. 
Hybrid mechanical systems enable new approaches to quantum control of mechanical objects, precision sensing, and quantum information processing.
}

\section{Introduction}
\label{Hybrid:sec:intro}

\index{hybrid system}\index{hybrid mechanical system}
The ability of functionalized mechanical systems to respond to electric, magnetic and optical forces has in the past led to widespread applications of mechanical resonators as sensitive force detectors. With improved technology the same principle will apply for resonators in the quantum regime and allow the integration of mechanical oscillators with a large variety of other (microscopic) quantum systems such as atoms and ions, electronic spins, or quantized charge degrees of freedom. The benefits of such hybrid quantum systems are quite diverse. \index{hybrid quantum system}\index{qubit}\index{cold atoms}\index{ions}\index{single electron spin}
On the one hand, the motion of the resonator can be used as a sensitive probe and readout device for static and dynamic properties of the quantum system. On the other hand, coupling the resonator to a coherent and fully controllable two-level system provides a way to prepare and detect non-classical states of mechanical motion. Finally, the mechanical system can serve as a quantum transducer to mediate interactions between physically quite distinct quantum systems. \index{quantum transducer}
This can be used to coherently couple e.g.\ an electronic spin to charge or optical degrees of freedom with various potential applications in the context of (hybrid) quantum information processing. \index{hybrid quantum information processing}

From a practical point of view the combination of mechanical resonators with microscopic quantum systems faces considerable challenges. Often the functionalization of mechanical resonators with electrodes, magnets, or mirrors competes with the requirement of a small mass to achieve a sufficient coupling strength on a single-quantum level. At the same time both the resonator and the other quantum system must be exceptionally well isolated from the environment to avoid decoherence. Various hybrid setups have been proposed that address those challenges, and some have already been implemented in experiments. This includes solid-state systems such as spin qubits, quantum dots, and superconducting devices, as well as atomic systems such as trapped atoms, ions, and molecules. In this chapter we give a brief overview of the different approaches towards mechanical hybrid quantum systems and discuss some basic examples from the fields of solid-state and atomic physics.



\section{Solid-State Quantum Systems Coupled to Mechanics}
\label{Hybrid:sec:solid}


Within the field of solid-state physics, a large variety of microscopic two- or few-level systems have been identified that are well isolated from the environment and allow for a coherent manipulation of their quantum state. Examples range from  electronic or nuclear spin states associated with naturally occurring defect centers \cite{Hybrid:Hanson2008} to electronic states of so-called artificial atoms such as quantum dots \cite{Hybrid:Shields2007} or superconducting Josephson devices \cite{Hybrid:Clarke2008}. Nanomechanical systems are naturally integrated with such solid-state quantum systems by fabricating them on the same chip, where they may interact with spins or charges via strong magnetic or electric forces. In contrast to most of the atomic implementations described below, the system dimensions are usually not limited by optical properties or trapping requirements, and without those restrictions strong interactions between an individual two-level system and a mechanical mode can be achieved more easily. 
On the other hand, it is more challenging to achieve long coherence times in the solid state.
In combination with cryogenic temperatures the solid-state approach to mechanical hybrid systems offers a promising route towards manipulating mechanical motion on a single-phonon level. 

We first present a brief overview of different physical mechanisms that have been suggested for achieving strong coupling between solid-state systems and mechanical motion. For the specific examples of superconducting charge qubits and electronic spin qubits we then describe mechanical sensing techniques and quantum control schemes as basic applications of these systems in the weak and strong interaction regime. 
\index{qubit}

\subsection{Overview of Systems and Coupling Mechanisms}\label{Hybrid:sec:overviewsolid}

The coupling of mechanical motion to other harmonic oscillators  has been  treated in other chapters of this book and we restrict the  discussion in this section
to microscopic two-level systems with a ground and excited state  $|g\rangle$ and $|e\rangle$. 
In solid-state systems the energy separation $E_{eg}$ between the two states is strongly dependent on the local electrostatic and magnetic fields, which also provides a way to couple to mechanical motion.  For example, by fabricating an oscillating electrode or a vibrating magnetic tip, the system energy $E_{eg}(\hat x)=E^0_{eg}+ \partial_x E_{eg} \hat x
+ (1/2) \partial^2_x E_{eg} \hat x^2+\dots$ now explicitly depends on the resonator displacement $\hat x=\xZPF (\bh+\bh^\dag)$. Even for nanoscale devices the zero point motion $\xZPF\approx 10^{-13}$ m is still much smaller than other system dimensions and corrections beyond the linear coupling term are usually negligible. Therefore, the generic Hamiltonian for the qubit-resonator system is given by \index{qubit}
\begin{equation}\label{Hybrid:eq:generic}
\hat H(t)= \hat H_q(t)+\hbar\OmegaM \bh^\dag \bh  + \hbar\lambda(\bh+\bh^\dag) \hat \sigma_z,
\end{equation}
where $\hat \sigma_z=|e\rangle\langle e| - |g\rangle\langle g|$ is the Pauli operator and $\hat H_q(t)$ denotes the unperturbed Hamiltonian for the solid-state qubit.  The relevant parameter in Eq.~(\ref{Hybrid:eq:generic}) is the coupling strength $\lambda=\partial_x E_{eg} \xZPF/2\hbar$,  which is the frequency shift per vibrational quantum. 

\begin{figure}[t]
\includegraphics[width=1\columnwidth]{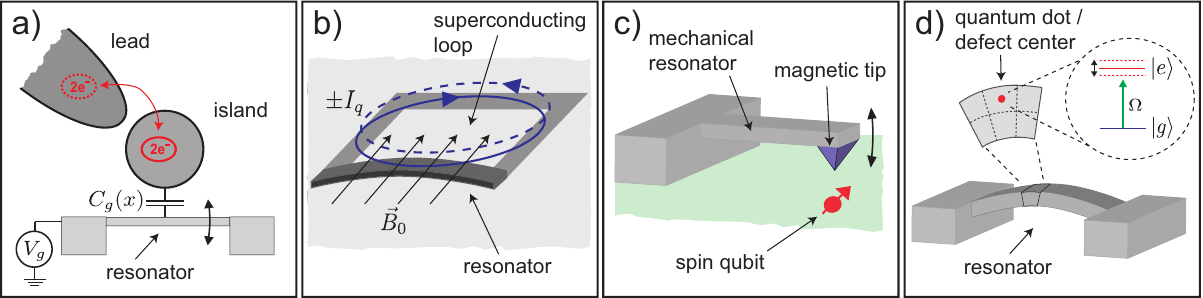}
\caption{Different schemes for coupling solid-state qubits and mechanical resonators. a) Electrostatic coupling to charge qubits. b) Lorentz force interactions with current states of a flux qubit.  c) Magnetic coupling to spins. d)  Deformation potential coupling to quantum dots or defect centers.
}
\label{Hybrid:fig:211}       
\end{figure}

While the basic form of the interaction in Eq.~(\ref{Hybrid:eq:generic}) has been derived from quite general  considerations, the origin and the magnitude of the qubit-resonator coupling $\lambda$ depends on the specific physical implementation. Figure~\ref{Hybrid:fig:211} illustrates four basic mechanisms for coupling different charge and spin qubits to mechanical motion. \index{charge qubit}\index{spin qubit}
In Fig.~\ref{Hybrid:fig:211}a two states encoded in quantized charge degrees of freedom, e.g. an electron on a quantum dot \cite{Hybrid:Blencowe2000,Hybrid:Knobel2003,Hybrid:Zippilli2009,Hybrid:Bennett2010} or a Cooper pair on a small superconducting island~\cite{Hybrid:Armour2002,Hybrid:Irish2003,Hybrid:Martin2004,Hybrid:LaHaye2009}, are coupled to a vibrating gate electrode. The energy for a total charge $Q$ on the island is $E_Q= (Q-Q_g)^2/2C_\Sigma$ where $C_\Sigma=C_0+C_g$ is the total capacitance of the island, $C_g$ the gate capacitance and $Q_g=V_gC_g$ the gate charge. For small displacements $C_g(x)\approx C_g(1- x/d)$ where $d$ is the gate separation, the typical coupling strength is
\begin{equation}\label{Hybrid:eq:lambdasuper}
\hbar \lambda_{el}\approx e V_g C_g/C_\Sigma\times \xZPF/d.
\end{equation}
For $d\approx 100$ nm and voltages up to $V_g=10$ V this coupling is quite substantial and can reach values in the range of $\lambda_{el}/2\pi\approx 5-50$ MHz~\cite{Hybrid:Armour2002}.

Instead of using charge states, a two-level system can alternatively be encoded in clockwise and anti-clockwise circulating currents in a superconducting loop~\cite{Hybrid:Zhou2006,Hybrid:Buks2006,Hybrid:Xue2007,Hybrid:Wang2008,Hybrid:Jaehne2008,Hybrid:Etaki2008} as shown in Fig.~\ref{Hybrid:fig:211}b. Here an interaction with a freely suspended arm of the loop can arise from the Lorentz force created by a magnetic field $B_0$ perpendicular to the bending motion~\cite{Hybrid:Xue2007,Hybrid:Wang2008,Hybrid:Jaehne2008}.  
For circulating currents of magnitude $I_q$  and a length $l$ of the resonator we obtain
\begin{equation}
\hbar \lambda_{Lor}\approx  B_0 I_q l \xZPF.
\end{equation}
Although the applied magnetic field is limited by the critical field of the superconductor, $B_0\leq10$ mT,  typical values of $I_q\approx 100$ nA and $l=5~\mu m$ still result in a coupling strength of $\lambda_{Lor}/2\pi\approx 0.1-1$ MHz~\cite{Hybrid:Xue2007,Hybrid:Wang2008,Hybrid:Jaehne2008}.

Qubits encoded in electronic or nuclear spin states can be coupled to the motion of a magnetized tip~\cite{Hybrid:Bargatin2003,Hybrid:Rugar2004,Hybrid:DegenTMV2009,Hybrid:Xue2007PRB,Hybrid:Rabl2009} as shown in Fig.~\ref{Hybrid:fig:211}c.  \index{electronic spin}\index{nuclear spin}
Here strong magnetic field gradients $\nabla B$ lead  to a position dependent Zeeman splitting of the spin states and for an electron spin,  
\begin{equation}\label{Hybrid:eq:lambdamag}
\hbar \lambda_{mag}\approx g_s\mu_B \xZPF \nabla B/2,
\end{equation}
where $\mu_B$ is the Bohr magneton and $g_s\approx 2$.  On the scale of a few nanometers, magnetic field gradients can be as high as  $\nabla B\sim 10^7$ T/m~\cite{Hybrid:Poggio2007,Hybrid:Mamin2012},Ê  which corresponds to a coupling strength of $\lambda_{mag}/2\pi\approx 10-100$ kHz~\cite{Hybrid:Rabl2009}.  Due to a smaller magnetic moment,  the coupling to a single nuclear spin is reduced by a factor $\sim10^{-3}$, but is partially compensated by the much longer coherence times of nuclear spin qubits.

Mechanical resonators cannot only modulate the configuration of externally applied fields, but for example also couple to quantum dots or defect centers by changing the local lattice configuration of the host material~\cite{Hybrid:WilsonRae2004}.  \index{quantum dots}
This deformation potential coupling is illustrated in Fig.~\ref{Hybrid:fig:211}d where flexural vibrations of the resonator induce a local stress  $\sigma \sim z_0\xZPF/l^2$, where $l$ is the resonator length and $z_0$ the distance of the defect from the middle of the beam. The corresponding level shift is given by 
\begin{equation}
\hbar \lambda_{def}\approx (D_e-D_g) z_0 \xZPF/l^2,
\end{equation}
where  $D_e$ and $D_g$ are deformation potential constants for the ground and excited electronic states. For quantum dots a coupling strength of $\lambda_{def}/2\pi\approx 1-10$ MHz can be achieved~\cite{Hybrid:WilsonRae2004}, but competes with radiative decay processes of the same order.

In summary, this brief overview shows that various different mechanisms lead to interactions between solid-state two-level systems and mechanical resonators. In many cases 
the single-phonon coupling strength $\lambda$ can be comparable to or even exceed the typical decoherence rate of the qubit $T_2^{-1}$ \index{decoherence} as well as the mechanical heating rate $\Gamma_\mathrm{th}=k_BT/\hbar Q$. As we describe now in more detail this enables various applications ranging from measurement and ground state cooling \index{ground state cooling} schemes for weak coupling to quantum control techniques in the strong coupling regime. \index{strong coupling regime}

\subsection{Superconducting Devices and Mechanics}

%

\index{superconducting devices}
Solid-state qubits which are encoded in a quantized charge degree of freedom can be coupled to mechanical motion via electrostatic interactions.  \index{qubit}
A prototype example  is the Cooper Pair Box (CPB), i.e. a  small superconducting island where Cooper pairs are coherently coupled to a large reservoir via a Josephson  tunnel junction (see Fig.~\ref{Hybrid:fig:211}a). \index{Cooper pair box}\index{Cooper pair}
The CPB belongs to a larger class of superconducting qubits~\cite{Hybrid:Makhlin2001,Hybrid:Clarke2008} and is  -- in its simplest realization --  described by the number $N$ of excess Cooper pairs on the island and its conjugate variable $\delta$ which is the difference of the superconducting phase across the Josephson junction. \index{Josephson junction}
As discussed in more detail in chapters [LEHNERT,CLELAND] the corresponding quantum operators obey the standard commutation relations  $[\hat N,\hat \delta]=i$  and the Hamiltonian operator for the CPB is
\begin{equation}\label{eq:FullHamiltonian}
\hat H_{\rm CPB} = E_C (\hat N-N_{g})^2-E_{J}\cos(\hat \delta).
\end{equation}
Here $E_C=4e^2/ 2C_\Sigma$  is the charging energy for a total island capacitance $C_\Sigma$, $E_J$ is the Josephson energy, and  $N_g=C_g V_g/(2e)$ the dimensionless gate charge  which can be adjusted by the voltage $V_g$ applied across the gate capacitance $C_g$. The CPB is usually operated in a regime where the charging energy $E_C/ h\sim 50$ GHz is 
the dominant energy scale
and the dynamics of the CPB is restricted to the two energetically lowest charge states. For example, by setting  $N_g =n+1/2 + \Delta N_g$ with integer $n$ and $\Delta N_g <1$,  these two states are $|g\rangle=|N=n\rangle$ and $|e\rangle = |N=n+1\rangle$ and form the basis of a so-called `charge qubit'~\cite{Hybrid:Makhlin2001}. The states $|g\rangle$ and $|e\rangle$ are separated by an adjustable charging energy $E_{eg}=2E_{C} \Delta N_g$ and  coupled by the Josephson tunneling term $\langle g|\hat H_{\rm CPB}|e\rangle = -E_J/2$.

When the gate electrode is replaced by a vibrating mechanical beam the capacitance $C_g(x)\approx C_g(1-x/d)$ varies with the beam displacement $x$ and the resulting  change in the charging energy $E_{eg}(x)$ introduces  a coupling between the qubit states and the mechanical resonator. The  Hamiltonian for the combined  system is  then given by~\cite{Hybrid:Armour2002,Hybrid:Irish2003}
\begin{equation} \label{Hybrid:eq:lambdasuperfull}
\hat H = E_{c} \Delta N_g \hat \sigma_{z} -  \frac{E_J}{2}\hat \sigma_x+ \hbar \OmegaM \bh^\dag \bh +\hbar \lambda (\bh+\bh^\dag)\hat \sigma_z\,,
\end{equation}
and we recover the general form of Eq.~(\ref{Hybrid:eq:generic}) of the qubit-resonator coupling with a single phonon coupling constant $\lambda\equiv\lambda_{el}$ as defined in Eq.~(\ref{Hybrid:eq:lambdasuper}). 
%
%
Due to the electrostatic nature of the interaction the achievable coupling strength $\lambda_{el}/2\pi\sim 10$ MHz between a charge qubit and a mechanical resonator can be substantially larger than the corresponding magnetic interactions with spin qubits discussed in section~\ref{Hybrid:sec:solidspin} below. However, for the same reason charge states are also more susceptible to random interactions with the environment and typical dephasing times $T_2$  for charge superposition states are in the pico- to nanosecond regime. 
An exception  to this rule occurs when the CPB is operated at the charge degeneracy point $\Delta N_g= 0 $. Here the eigenstates of $\hat H_{\rm CPB}$, namely   $|\tilde g \rangle=(|g\rangle +| e \rangle)/\sqrt{2}$ and $|\tilde e \rangle=(|g\rangle -| e\rangle)/\sqrt{2}$,  are combinations of different charge states and therefore highly insensitive to ubiquitous sources of low frequency electric noise~\cite{Hybrid:Vion2002}. By assuming $\Delta N_g=0$ and re-expressing Eq.~(\ref{Hybrid:eq:lambdasuperfull}) in terms of the Pauli operators  $\tilde \sigma_j$ for the rotated basis states $|\tilde g \rangle$, $|\tilde e \rangle$  we obtain 
\begin{equation} \label{Hybrid:eq:lambdasuperrot}
\hat H =  \frac{E_J}{2}\tilde \sigma_z+ \hbar \OmegaM \bh^\dag \bh +\hbar \lambda (\bh+\bh^\dag)\tilde \sigma_x\,,
\end{equation}
as our final model for the coupled resonator charge-qubit system. Indeed, by using optimized charge qubit designs dephasing times $T_2 >1\, \mu$s have been demonstrated~\cite{Hybrid:Vion2002,Hybrid:Schreier2008}. \index{charge qubit}
This makes the CPB a promising candidate to achieve strong coupling $\lambda T_2 >1$ with a mechanical resonator.

Equation~(\ref{Hybrid:eq:lambdasuperrot}) is familiar from related models studied in the context of cavity QED~\cite{Hybrid:Raimond2001} or trapped ions~\cite{Hybrid:Leibfried2003}, where usually a resonant exchange of excitations between the qubit and the resonator is used for cooling or quantum control of the resonator mode. \index{cavity quantum electrodynamics}
However, with the exception of the high frequency dilatation modes described in chapter [CLELAND], mechanical frequencies of $\mu$m sized beams are typically in the range of  $10-100$ MHz  and Eq.~(\ref{Hybrid:eq:lambdasuperrot}) describes a highly non-resonant coupling to a qubit with a transition frequency $E_J/h\sim 5$ GHz. 
Therefore, in the following we briefly outline two possibly strategies for potential quantum applications in the present  system.  First, we remark that in the relevant regime  $\lambda \ll \OmegaM < E_J$ second order perturbation theory can be used to approximate Eq.~(\ref{Hybrid:eq:lambdasuperrot}) by an effective Hamiltonian~\cite{Hybrid:Irish2003,Hybrid:LaHaye2009} 
\begin{equation}
\hat H\simeq  \frac{E_J}{2}\tilde \sigma_z+ \hbar \OmegaM \bh^\dag \bh +\hbar  \chi ( \bh^\dag \bh +1/2) \tilde \sigma_z\,.
\end{equation}
The resulting coupling term with a strength $\hbar \chi= (\hbar\lambda)^2\times 2 E_J/(E_J^2-(\hbar \OmegaM)^2)$ can be interpreted as a shift of the qubit frequency proportional to the phonon number $\bh^\dag \bh$. Under the condition $\chi T_2 > 1$ this frequency shift can in principle be detected and the charge qubit can be used to implement a quantum non-demolition measurement  of the number of vibrational quanta of the mechanical mode. To resolve a single vibrational level in time the coupling $\chi$ must also exceed the rate $\Gamma_\mathrm{th}$ at which the environment induces jumps between different vibrational states of the resonator. Estimates show that in this setting the combined condition $\chi > \Gamma_\mathrm{th},T_2^{-1}$ for a phonon resolved measurement is experimentally feasible~\cite{Hybrid:Irish2003}.
  
To go beyond passive measurement applications a second strategy is to realize  effective resonance conditions by applying an oscillating gate voltage $V_g(t)\sim \cos(\omega_0 t)$ such that the microwave frequency $\omega_0= E_J/\hbar - \OmegaM$ is used to gap the energy between vibrational and qubit excitations~\cite{Hybrid:Martin2004,Hybrid:Rabl2004}.  
In the interaction picture with respect to the free evolution $\hat H_0=E_J/2 \tilde \sigma_z + \hbar \OmegaM \bh^\dag \bh$ the resulting Hamiltonian is then of the form 
\begin{equation}\label{Hybrid:eq:supercondJC}
\hat H\simeq 
\lambda (\tilde \sigma_+ \bh + \tilde \sigma_- \bh^\dag)  
+ \mathcal{O}\left(e^{\pm i2\OmegaM t}, e^{\pm i 2\omega_0 t}\right), 
\end{equation}
where for $\lambda\ll \OmegaM,\omega_0$ the oscillating terms can be neglected by using a rotating wave approximation. Hamiltonian (\ref{Hybrid:eq:supercondJC}) reduces to the resonant  Jaynes-Cummings model which allows a coherent exchange of qubit and vibrational excitation. \index{Jaynes-Cummings model}
Applications of this model such as sideband cooling, state preparation and detection have been discussed in different areas of quantum optics \cite{Hybrid:Haroche2006}.  The driven CPB provides the tool to implement similar applications~\cite{Hybrid:Martin2004,Hybrid:Rabl2004} for the vibrational modes of macroscopic mechanical resonators. 
We close this section by noting that strong coupling of a superconducting phase qubit to a mechanical oscillator has been observed experimentally \cite{Hybrid:OConnell2010}, as discussed in more detail in chapter [CLELAND].

\subsection{Spin Qubits and Mechanics}\label{Hybrid:sec:solidspin}

\index{spin qubit}\index{electronic spin}\index{nuclear spin}\index{qubit}
As discussed in section~\ref{Hybrid:sec:overviewsolid}, electronic and nuclear spin states can be
coupled to mechanical motion by way of a magnetic field gradient.  In
the solid state, this situation is realized by positioning a spin
qubit -- typically residing in the lattice of some material -- in
close proximity to a strongly magnetized tip.  One of the two
elements, the tip or the qubit, is then rigidly affixed to a
cantilever or other mechanical resonator.  The most prominent examples
of such experiments include mechanically detected magnetic resonance
and optical experiments on nitrogen-vacancy defects in diamond.

\subsubsection{Mechanical Detection of a Single Electron Spin}

\index{single electron spin}
The first experiments demonstrating coupling between a nanomechanical
cantilever \index{cantilever} and the spin of an isolated single electron appeared in
2004.  In a landmark experiment, Rugar \textit{et al.}\ measured the
force of flipping a single unpaired electron spin contained in a
silicon dangling bond (commonly known as an E' center) using a NEMS
cantilever \cite{Hybrid:Rugar2004}.  This achievement concluded a
decade of development of a technique known as magnetic resonance force
microscopy (MRFM) and stands out as one of the first single-spin
measurements in a solid-state system. \index{magnetic resonance force microscopy}\index{MRFM}\index{single-spin detection}

\begin{figure}
\sidecaption[t]
\includegraphics[width=0.5\columnwidth]{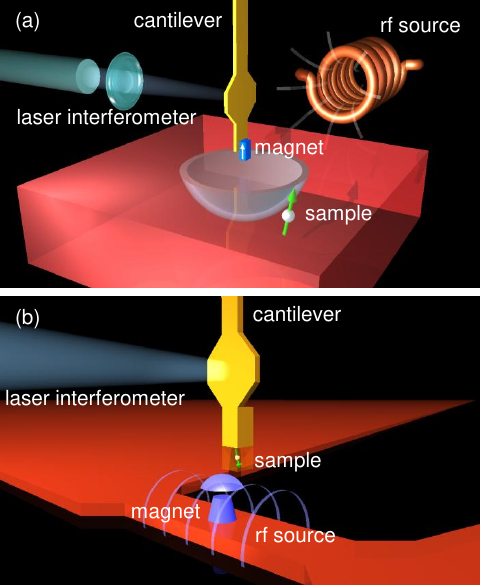}
\caption{Schematics of an MRFM apparatus.  (a) ``Tip-on-cantilever'' arrangement, \index{cantilever} such as used
    in the single electron MRFM experiment of 2004
    \cite{Hybrid:Rugar2004}.  (b)   ``Sample-on-cantilever'' arrangement, like the one used for the
    nanoscale virus imaging experiment in 2009
    \cite{Hybrid:DegenTMV2009}. In both cases the hemispherical region
    around the magnetic tip is the region where the spin resonance
    condition is met -- the so-called ``resonant slice''.}
\label{Hybrid:fig:221}
\end{figure}

The principle behind MRFM is simple (see Fig.~\ref{Hybrid:fig:221}).  Magnetic moments -- such as
those associated with single electron or single nuclear spins --
produce a force when in a magnetic field gradient: $F = \mu \nabla B$, where $\mu$ is the spin's magnetic moment and $\nabla B$ is the spatial field gradient.  This force can couple
the deflection of a compliant cantilever to the spin if either the spin
or the gradient source, such as a small magnet, are fixed to the
cantilever.  If the magnetic field gradient is large enough (i.e.\ the
coupling is strong enough), the spin polarization and the cantilever's
motion will be coupled.

Most MRFM techniques utilize this coupling to make measurements of
spin density on the micro- or nanometer scale
\cite{Hybrid:Poggio2010}.  They employ extremely compliant cantilevers \index{cantilever}
capable of detecting forces as small as $1~\mathrm{aN}/\sqrt{\mathrm{Hz}}$ and optical
interferometers that can measure the cantilever's displacement to
resolutions of $1~\mathrm{pm}/\sqrt{\mathrm{Hz}}$.  Furthermore, magnetic resonance
techniques are used in order to selectively address ensembles of spins
in a sample, allowing for both spatial and chemical selectivity.  
For example, a pulse sequence known as a rapid adiabatic passage can be
applied using a radio-frequency (rf) source.  
Each adiabatic passage pulse causes only the spins on resonance with the rf carrier frequency to flip. Using a periodic pulse sequence, a small ensemble of spins -- at a particular region in space -- can be made to flip at nearly any desired frequency.  By choosing the flip frequency to be at the cantilever's mechanical resonance, the periodic spin oscillations will in turn drive the cantilever into oscillation.
Because this spin force is made to occur at the cantilever's resonant
frequency and at a particular phase, it can be distinguished from all
other random electro-static and thermal forces disturbing the
cantilever's motion.

We should note that the force sensitivity required for the single
electron spin measurement imposed several limitations.  Since the
single-shot signal-to-noise ratio (SNR) was 0.06, up to 12 hours of
averaging per data point were required \cite{Hybrid:Rugar2004}.  In addition to making useful
imaging prohibitively time-consuming, this small SNR precludes the
technique from following the dynamics of the single electron spin. \index{single electron spin}
Large SNRs would allow for shorter measurement times.  If the
measurement time could be reduced below the correlation time of the
electron spin, which is related to the relaxation time of the electron
spin in its rotating frame, real-time readout of the spin quantum
state would become possible.  Such readout would enable a wide variety
of quantum measurement experiments.  In the case of the electron spin
E' centers, the correlation time was measured to be 760 ms.  Until the
SNR improves dramatically, real-time readout of spins will only be
possible for small ensembles of electrons rather than for single
electrons.  One example is the real-time measurement of the direction
of spin polarization for an ensemble of spins ($\sim 70$ electron
spins) \cite{Hybrid:Budakian2005}.  In contrast, due to low SNR, the
single-spin measurement \cite{Hybrid:Rugar2004} could not discern the direction of the
measured spin; it could only ascertain its position.

\subsubsection{Mechanical Detection of Nuclear Spins}

\index{nuclear spin}\index{nuclear spins}
Using this technique to couple and detect a single nuclear spin is far
more challenging than to detect a single electron spin.  The magnetic
moment of a nucleus is much smaller than that of an electron: a $^1$H
nucleus (proton), for example, possesses a magnetic moment that is only
$\sim 1/650$ of an electron spin moment.  Since the measured force,
$F$, is directly proportional to the magnetic moment of the spin, a
significantly higher force resolution is required for nuclear spin
experiments than for electron spin experiments.  Other important
nuclei, such as $^{13}$C or a variety of isotopes present in
semiconductors, have even weaker magnetic moments than $^1$H.  In
order to observe single nuclear spins, it is necessary to improve the
state-of-the-art sensitivity by another two to three orders of
magnitude.  While not out of the question, this is a daunting task
that requires significant advances to all aspects of the MRFM
technique.

\subsubsection{Strong Magnetic Coupling}

\index{magnetic coupling}
The strong magnetic coupling achieved between spins and the cantilever
enables the high sensitivity of MRFM.  This coupling is mediated by
field gradients that can exceed $5 \times 10^6$~T/m
\cite{Hybrid:Poggio2007,Hybrid:Mamin2012}.  For the cantilevers and
magnetic tips used in these experiments this corresponds to
$\lambda_{mag} / 2 \pi \approx 10$ kHz for a single
electron spin and $10$ Hz for a proton.  Such high
gradients have been achieved using micro-fabricated Dy or FeCo
magnetic tips and by the ability to make stable measurements with the
sample positioned less than 50~nm from the apex of the tip.  The
strong interaction between spins and the mechanical sensor has been the
subject of a number of theoretical studies, and is predicted to lead
to a host of intriguing effects.  These range from shortening of spin
lifetimes by ``back action''
\cite{Hybrid:Mozyrsky2003,Hybrid:Berman2003}, to spin alignment by
specific mechanical modes either at the Larmor frequency or in the
rotating frame \cite{Hybrid:Mangusin2003}, to a mechanical analog of a laser \cite{Hybrid:Bargatin2003}, and to long-range mediation of
spin couplings using charged resonator arrays \cite{Hybrid:Rabl2010}.

The first direct experimental evidence for accelerated nuclear spin
relaxation induced by a single, low-frequency mechanical mode was
reported in 2008 \cite{Hybrid:DegenPRL2008}.  In these experiments the
slight thermal vibration of the cantilever generated enough magnetic
noise to destabilize the spin.  Enhanced relaxation was found when one
of the cantilever's upper modes (in particular the third mode with a
frequency of about 120 kHz) coincided with the Rabi frequency of the
$^{19}$F spins in CaF$_2$.  In this regime, the
spins are more tightly coupled to one mechanical resonator mode than
to the continuum of phonons that are normally responsible for
spin-lattice relaxation.  Interestingly, these initial experiments
showed a scaling behavior of the spin relaxation rate with important
parameters, including magnetic field gradient and temperature, that is
substantially smaller than predicted by theory.

\subsubsection{Nano-MRI and Potential Practical Applications}

The coupling of small nuclear spin ensembles to a compliant mechanical oscillator through strong magnetic tips has resulted in the highest magnetic resonance imaging (MRI) resolution achieved by any method.  \index{magnetic resonance imaging}
In 2009, Degen
\textit{et al.}\ demonstrated three-dimensional (3D) MRI of $^1$H
nuclear spins \index{nuclear spins} in a biological specimen (tobacco mosaic virus) with a
spatial resolution down below 10 nm \cite{Hybrid:DegenTMV2009}.  This
resolution represents a 100 million-fold improvement in volume
resolution over conventional MRI and shows the potential of MRFM as a
tool for elementally selective imaging on the nanometer scale.  If the
development of such techniques continues, these results indicate that
force-detected spin resonance has the potential to become a
significant tool for structural biologists.

\subsubsection{Nitrogen Vacancy Centers and Mechanics}

\index{nitrogen vacancy center}\index{NV center}
Recently, single nitrogen vacancy (NV) centers hosted in diamond have
been proposed as solid-state qubits amenable to mechanical coupling \cite{Hybrid:Rabl2009}.
Again, as in the case of MRFM, the coupling rests on bringing the
qubit -- in this case a single NV -- in close proximity to a strongly
magnetized tip.  Then, either the NV or the tip must be affixed to a
mechanical oscillator.  NV centers appear especially attractive qubits
due to their excellent optical and electronic properties.  A single NV
spin can be readily initialized and measured by optical means,
manipulated using resonant rf pulses, and excellent coherence times
up to a few milliseconds persist even in ambient conditions
\cite{Hybrid:Jelezko2004,Hybrid:BalasubramanianNatMat2009}.  As a
result NVs have been proposed and used as ultra-sensitive scanning
magnetic
sensors \cite{Hybrid:Degen2008,Hybrid:Maze2008,Hybrid:BalasubramanianNat2009,Hybrid:Grinolds2011}.

First experiments have recently demonstrated the
coupling of the NV spin to mechanical motion.  Arcizet \textit{et
  al.}\ coupled an NV to the motion of a SiC nanowire using field gradients around $7 \times 10^3$ T/m  \cite{Hybrid:Arcizet2011}. The NV was fixed to the tip of the nanowire while the magnet was placed near to it. Nanowire vibrations of a few tens of nanometers in amplitude were detected through a change in the lineshape of the NV spin resonance. 
 More recently, Kolkowitz \textit{et al.}\ used an NV spin to sense the vibrations of a cantilever resonator \index{cantilever} with a magnetic tip \cite{Hybrid:Kolkowitz2012}. A sequence of coherent manipulation pulses was applied to the spin in order to enhance its sensitivity to the resonator vibrations while suppressing noise from other sources. In this way, mechanical vibrations down to a few picometers in amplitude were detected without phase locking NV spin dynamics and resonator vibrations.
In these initial experiments, the spin-resonator coupling strength was  $\lambda_{mag} / 2 \pi = 70$~Hz \cite{Hybrid:Arcizet2011} or lower \cite{Hybrid:Kolkowitz2012}. Coupling strengths in the kHz range could be reached by combining a strong magnet with a nanoscale oscillator with large zero-point motion \cite{Hybrid:Kolkowitz2012}.

\subsubsection{Increasing the magnetic coupling strength}

The prospects of improving hybrid mechanical systems based on spin qubits depend
on progress in increasing the magnetic coupling strength
$\lambda_{mag}$.  First, the magnetic tips can and must be improved
with the use of cleaner materials and lithographic processing
techniques.  Second, the development of experimental techniques
designed to bring the spin qubit and the gradient source as close
together as possible without destroying either qubit coherence or
introducing mechanical dissipation should also yield significant gains
in coupling strength.

\section{Atoms, Ions, and Molecules Coupled to Mechanics}
\label{Hybrid:sec:atoms}

Atoms, ions, and molecules are quantum systems par excellence, and a sophisticated toolbox exists for coherent manipulation of their electronic, spin, and motional degrees of freedom \cite{Hybrid:Chu2002,Hybrid:Kasevich2002}. It is therefore natural to ask whether such atomic quantum systems can be coupled to mechanical oscillators. Through the coupling, the tools of atomic physics could become available for quantum control of mechanical devices. On the other hand, mechanical oscillators could find new applications in atomic physics experiments, such as optical lattices with vibrating mirrors. \index{atoms}\index{ultracold atoms}\index{cold atoms} \index{ions}

Compared to the solid-state based approaches discussed in the previous sections, coupling atomic systems to mechanical oscillators creates a qualitatively different setting. Atoms in a trap can be regarded as mechanical oscillators themselves. Acting as a dispersive medium inside an optical cavity, an interesting variant of cavity optomechanics in the quantum regime can be realized (see chapter [STAMPER-KURN]). 
Atomic systems offer both the continuous degree of freedom of their motion in a trap as well as a discrete set of internal electronic and spin states that can be reduced to two-level systems. Both the internal and motional state can be initialized, coherently manipulated, and detected on the quantum level with high fidelity, using techniques that have been developed 
in experiments on atomic clocks and interferometers \cite{Hybrid:Wynands2005,Hybrid:Cronin2009}, Bose-Einstein condensation \cite{Hybrid:Ketterle1999,Hybrid:Bloch2008,Hybrid:Reichel2011}, and quantum information processing \cite{Hybrid:Wineland2009}. \index{quantum information processing} \index{Bose Einstein condensate}
Coherence times of atomic systems are typically in the range of milliseconds up to many seconds \cite{Hybrid:Deutsch2010}, and thus much longer than those of most solid-state quantum systems discussed in the previous sections. Moreover, many properties of  atomic systems can be widely tuned in-situ with external fields, including trapping frequencies, laser cooling rates, and even the strength of atom-atom interactions. 

While the good isolation of atoms trapped in a vacuum chamber enables long coherence times, it renders  coupling to mechanical oscillators more challenging.
Various coupling mechanisms have been proposed, such as electrostatic coupling to the motion of trapped ions \cite{Hybrid:Wineland1998,Hybrid:Tian2004,Hybrid:Hensinger2005} and molecules \cite{Hybrid:Singh2008}, magnetic coupling to atomic spins \cite{Hybrid:Treutlein2007,Hybrid:Geraci2009,Hybrid:Singh2010,Hybrid:Joshi2010,Hybrid:Steinke2011}, and optomechanical coupling to atoms in free space \cite{Hybrid:Hammerer2009a,Hybrid:Hammerer2010,Hybrid:Vogell2013} and in optical cavities \cite{Hybrid:Meiser2006,Hybrid:Genes2008,Hybrid:Ian2008,Hybrid:Hammerer2009b,Hybrid:Wallquist2010,Hybrid:Bhattacherjee2009,Hybrid:Genes2009,Hybrid:Zhang2010,Hybrid:Paternostro2010,Hybrid:DeChiara2011,Hybrid:Chang2011}. Remarkably, some of these schemes predict strong atom-oscillator coupling even for a single atom \cite{Hybrid:Hammerer2009b,Hybrid:Wallquist2010}. First experimental implementations of hybrid atom-oscillator systems have recently been reported \cite{Hybrid:Wang2006,Hybrid:Hunger2010,Hybrid:Camerer2011,Hybrid:Korppi2013}. In the following, we discuss several of these proposals and experiments (see also the review in \cite{Hybrid:Hunger2011}).

\subsection{Direct Mechanical Coupling} \label{Hybrid:sec:direct}

The conceptually most straightforward approach is to directly couple the vibrations of a mechanical oscillator to the vibrations of an atom or ion in a trap with the help of a ``spring'', i.e.\ a distance-dependent force between the two systems \cite{Hybrid:Hunger2011}. 
We consider an atom of mass $m_{at}$ in a trap of frequency $\Omega_{at}$ and a mechanical oscillator of mass $\mass$ and frequency $\OmegaM$. The coupling force derives from a potential $U_c(d)$ that depends on the distance $d$ between the oscillator and the atom. 
For small displacements $\hat x, \hat x_{at} \ll d$, the resulting coupling Hamiltonian is of the form $\hat H_c=U_c''(d) \hat x \, \hat x_{at} $, where $U_c''(d)$ is the curvature of $U_c$ evaluated at the mean atom-oscillator distance $d$. The oscillator displacement can be written as $\hat x = \xZPF(\bh+\bh^\dagger)$ in terms of creation/annihilation operators $[\bh,\bh^\dagger]=1$ and the zero-point amplitude $\xZPF=\sqrt{\hbar/2\mass\OmegaM}$. Similarly, the atomic displacement is $\hat x_{at}= x_{at,0} (\ah_{at}+\ah_{at}^\dagger)$ with $[\ah_{at},\ah_{at}^\dagger]=1$  and $x_{at,0}=\sqrt{\hbar/2 m_{at} \Omega_{at}}$. The Hamiltonian of the coupled system is then given by 
\begin{equation}
\hat H = \hbar\Omega_{at} \ah^{\dagger}_{at}\ah_{at} + \hbar\OmegaM \bh^{\dagger}\bh + \hbar\lambda (\ah_{at} + \ah^{\dagger}_{at})(\bh + \bh^{\dagger}),
\end{equation}
with a single-phonon atom-oscillator coupling constant
\begin{equation}\label{Hybrid:Eq:directcoupling}
\hbar\lambda=U_c''(d) \xZPF \, x_{at,0} \simeq \epsilon \frac{\hbar\Omega_{at}}{2}\sqrt{\frac{m_{at}}{\mass}},
\end{equation}
for near-resonant coupling $\Omega_{at}\simeq \OmegaM$. It is important to note that $U_c$ also modifies the atomic trapping potential by contributing a term of order $U_c''(d)\hat x_{at}^2$ \cite{Hybrid:Hunger2011}. 
We therefore introduce the dimensionless parameter $\epsilon = U_c''(d) / (m_{at} \Omega_{at}^2)$, which compares $U_c''(d)$ to the curvature of the atom trap. To avoid strong trap distortion, we typically have $\epsilon \ll 1$. In the special case where $U_c$ itself provides the atom trap, we have  $\epsilon = 1$. To achieve $\epsilon>1$, the effect of $U_c$ on the trap has to be partially compensated, requiring sophisticated trap engineering. For direct mechanical coupling, we thus find that $\lambda$ scales with $\Omega_{at}$ but is reduced by the atom-oscillator mass ratio, which is typically very small ($\sqrt{m_{at}/\mass}\sim 10^{-8} - 10^{-4}$). To achieve significant coupling strength, oscillators with small $\mass$ are advantageous. 

Several theoretical proposals consider direct mechanical coupling between a trapped ion and an oscillator with a metallic electrode on its tip \cite{Hybrid:Wineland1998,Hybrid:Tian2004,Hybrid:Hensinger2005}. \index{ion trap} \index{trapped ion}\index{ions}
In this case, $U_c = eq/(4\pi\epsilon_0 d)$ is the Coulomb interaction between the ion of elementary charge $e$ and the charge $q=C_qV_q$ on the oscillator tip. For a nanoscale oscillator with $\mass = 10^{-15}$~kg coupled to a single $^9$Be$^+$ ion in a trap with $\Omega_{at}/2\pi = 70$~MHz, we obtain $\lambda/2\pi = \epsilon \times 150$~Hz assuming $\OmegaM = \Omega_{at}$  \cite{Hybrid:Hunger2011}. A value of $\epsilon = 1$ can be achieved e.g.\ with $d=10~\mu$m and a metallic tip with a capacitance $C_q=10^{-17}$~F and an applied voltage $V_q = 90$~V.

Stronger coupling is possible if $N\gg 1$ atoms are simultaneously coupled to the mechanical oscillator. In this case, $\ah_{at}$ and $\ah_{at}^\dagger$ refer to the atomic center-of-mass (COM) motion. The coupling is collectively enhanced by a factor $\sqrt{N}$, so that
\begin{equation}\label{Hybrid:Eq:directcouplingCOM}
\lambda_N=\lambda\sqrt{N}=\epsilon \frac{\Omega_{at}}{2}\sqrt{\frac{N m_{at}}{\mass}} .
\end{equation} \index{collective coupling}\index{collective enhancement}
This result  can be intuitively obtained by replacing $m_{at} \rightarrow N m_{at}$ in Eq.~(\ref{Hybrid:Eq:directcoupling}), for a derivation see \cite{Hybrid:Hammerer2010}. 
An example of such collective coupling where $\lambda_N$ can reach several kHz is given in section~\ref{Hybrid:sec:latticecoupling}.

\begin{figure}[b]
\sidecaption
\includegraphics[width=7.5cm]{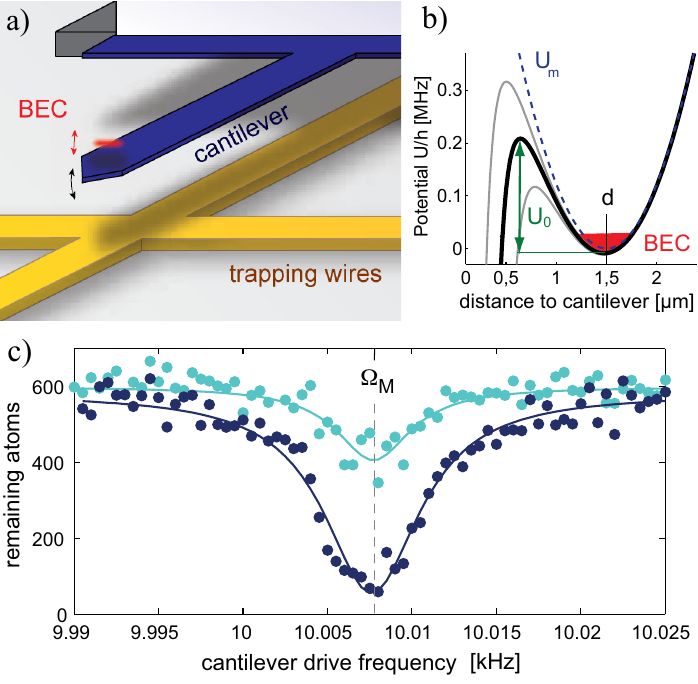}
\caption{Coupling of a mechanical cantilever \index{cantilever} and an atomic BEC via atom-surface forces \cite{Hybrid:Hunger2010}. (a) Atom chip with cantilever oscillator (length 200~$\mu$m, $\OmegaM /2\pi = 10$~kHz, $\mass = 5$~ng, $Q=3200$). \index{atom chip} The atoms can be trapped and positioned near the cantilever with magnetic fields from wire currents. (b) Combined magnetic trapping and surface potential. The surface potential reduces the trap depth to $U_0$. Cantilever oscillations modulate the potential, thereby coupling to atomic motion. (c) Cantilever resonance detected with the atoms, for two different driving strengths of the cantilever. }
\label{Hybrid:fig:BECcantilever}       
\end{figure}

In the experiment of \cite{Hybrid:Hunger2010},  a direct mechanical coupling between a  cantilever oscillator \index{cantilever} and ultracold atoms was demonstrated for the first time. \index{cold atoms}
An atomic Bose-Einstein condensate (BEC) of $N = 2 \times 10^3$ atoms was placed at about one micrometer distance from the surface of the cantilever and used as a probe for cantilever oscillations (see Fig.~\ref{Hybrid:fig:BECcantilever}a). \index{Bose Einstein condensate} \index{atom trap} \index{trapped atom}
The coupling potential $U_c$ is due to attractive atom-surface interactions, which substantially modify the magnetic trapping potential $U_m$ at such small distance. One effect of the surface force is to reduce the potential depth (see Fig.~\ref{Hybrid:fig:BECcantilever}b). In addition, it shifts the trap frequency and minimum position. When the cantilever oscillates, the trapping potential is modulated at the cantilever frequency $\OmegaM$, resulting in mechanical coupling to the atoms as described above.

In the experiment, the vibrating cantilever induced large-amplitude atomic motion that was detected simply via atom loss across the barrier $U_0$, see Fig.~\ref{Hybrid:fig:BECcantilever}c. The observed atom-cantilever coupling depends strongly on the trap parameters and shows resonant behavior if $\OmegaM = \Omega_{at}$. Coupling to collective mechanical modes of the BEC other than the COM mode was observed as well \cite{Hybrid:Hunger2010}. 
While it was possible to detect the cantilever motion with the atoms, the backaction of atoms onto the cantilever was negligible in this experiment, mainly because the relatively large $\mass = 5$~ng results in a small coupling constant $\lambda_N/2\pi \simeq 10^{-2}$~Hz. Much stronger coupling could be achieved by miniaturizing the cantilever. Since coupling via surface forces does not require functionalization of the cantilever, it could be used to couple atoms to molecular-scale oscillators such as carbon nanotubes \cite{Hybrid:Gierling2011}. In this case, a coupling constant of a few hundred Hz could be achieved \cite{Hybrid:Hunger2011}.

\subsection{Magnetic Coupling to Atomic Spin}

\index{magnetic coupling}\index{atomic spin}
The vibrations of a mechanical oscillator can also be coupled to the spin of the atoms. This has several advantages compared to coupling to atomic motion. First, the atomic spin can be manipulated with higher fidelity. For hyperfine spins, coherence times $T_2$ of many seconds have been achieved \cite{Hybrid:Deutsch2010}. Second, it is easier to isolate a two-level system among the internal states, providing a way to the preparation of non-classical quantum states of the oscillator. Third, hyperfine spin transition frequencies lie in the MHz to GHz range,  significantly higher than typical trap frequencies. This enables coupling to high-frequency mechanical oscillators, which are easier to cool to the ground state.

\begin{figure}
\centering
\includegraphics[width=0.9\columnwidth]{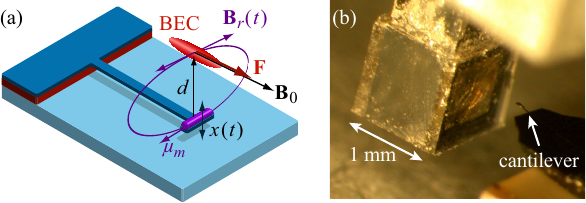}
\caption{(a) Schematic setup for coupling the spin of an atomic BEC to a nanoscale cantilever \index{cantilever} with a magnetic tip \cite{Hybrid:Treutlein2007}. \index{Bose Einstein condensate} Cantilever oscillations $x(t)$ are transduced by the magnet into an oscillating  field $B_r(t)$, which couples to the spin $\mathbf{F}$ of the atoms. (b) Experimental setup of \cite{Hybrid:Wang2006}, where a vibrating cantilever with a magnetic tip induces atomic spin resonance in a room-temperature Rb vapor cell. }
\label{Hybrid:fig:BECspin}       
\end{figure}

To couple to the spin of the atoms, the oscillator is functionalized with a small magnet that generates a field gradient $\nabla B$, see Fig.~\ref{Hybrid:fig:BECspin}a. Mechanical oscillations $x(t)$ are transduced into an oscillating magnetic field $B_r(t) = \nabla B \, x(t)$ that couples to the spin. If $B_r(t)$ is  perpendicular to the static field $B_0$ in the trap, the  Hamiltonian is 
\begin{equation}
\hat H = \frac{\hbar \omega_L}{2}  \hat \sigma_z + \hbar\OmegaM \bh^\dag \bh  + \hbar\lambda_{mag}(\bh+\bh^\dag) \hat \sigma_x
\end{equation}
with a coupling constant $\lambda_{mag} = g \mu_B \xZPF \nabla B / 2\hbar$ similar to Eq.~(\ref{Hybrid:eq:lambdamag}). 
The prefactor $g$ (of order unity) accounts for the matrix element of the  atomic hyperfine transition considered. For transitions between Zeeman sublevels, the Larmor frequency is  $\omega_L = g_F \mu_B B_0/\hbar$, with the hyperfine Land{\'e} factor $g_F$. It can be widely tuned by adjusting $B_0$ in order to achieve resonance $\omega_L = \OmegaM$.
The coupled system realizes a mechanical analog of the Jaynes-Cummings model in cavity quantum electrodynamics \cite{Hybrid:Raimond2001}, with the \emph{phonons} of the mechanical oscillator playing the role of the \emph{photons} of the electromagnetic field. \index{cavity quantum electrodynamics} \index{Jaynes-Cummings model}

To achieve large $\lambda_{mag}$ a strong gradient $\nabla B$ is required. Approximating the magnet by a dipole of magnetic moment $\mu_m$, we have $\nabla B = 3\mu_0\mu_m/4\pi d^4$. It is thus essential to trap and position the atoms at very small distance $d$ from the oscillator tip. At the same time, care has to be taken that $\nabla B$ does not significantly distort the atomic trapping potential. With neutral atoms in magnetic microtraps, $d$ can be as small as a few hundred nanometers \cite{Hybrid:Hunger2010}. Compared with the solid-state implementations discussed in section \ref{Hybrid:sec:solidspin}, where $d$ can be in the tens of nanometers range, it is thus more difficult to achieve large $\nabla B$. On the other hand, the spin decoherence rates $T_2^{-1}$ of trapped atoms are exceptionally small. The main challenge is the thermal decoherence \index{decoherence} rate $\Gamma_\mathrm{th}=k_B T /\hbar Q$ of the mechanical oscillator. 
The single-phonon single-atom strong coupling regime requires $\lambda_{mag}>\Gamma_\mathrm{th},T_2^{-1}$. \index{strong coupling regime}
As in the previous section, collective coupling to $N\gg 1$ atoms is a possible strategy to enhance the coupling strength. In this case, the collective strong-coupling regime requires $\lambda_N = \lambda_{mag}\sqrt{N}>\Gamma_\mathrm{th},T_2^{-1}$. \index{collective enhancement} \index{collective coupling} 
Several theory papers investigate how to achieve the strong coupling regime by coupling nanoscale cantilevers to the spin of ultracold neutral atoms \cite{Hybrid:Treutlein2007,Hybrid:Geraci2009,Hybrid:Singh2010,Hybrid:Joshi2010,Hybrid:Steinke2011}. The predicted coupling constants lie in the range of $\lambda/2\pi \approx 10-10^3$ Hz, so that very high $Q$ and low $T$ are required for strong coupling.


In \cite{Hybrid:Wang2006}, a first experiment was reported where a cantilever with a magnetic tip was coupled to atoms in a room-temperature vapor cell, see Fig.~\ref{Hybrid:fig:BECspin}b. The cantilever was piezo-driven and induced spin resonance in the atomic vapor, which was recorded with a laser. Besides demonstrating spin-oscillator coupling, such a setup is of interest for applications in magnetic field sensing, where the cantilever is essentially used as a tool for spectroscopy of the atomic transition frequency.

\subsection{Optomechanical Coupling in Free Space}\label{Hybrid:sec:latticecoupling}

\index{optomechanical coupling}
The coupling mechanisms discussed in the preceding sections require to position atoms close to the mechanical oscillator. Combining trapping and cooling of atoms in ultra-high vacuum (UHV) with a cryogenic environment as required for minimizing decoherence of the micromechanical system is a demanding task. In contrast, an indirect coupling mechanism acting over some distance would allow to keep the atomic and the micromechanical system in separate environments. Such a scheme was suggested in \cite{Hybrid:Hammerer2010,Hybrid:Vogell2013} and experimentally implemented as described in \cite{Hybrid:Camerer2011,Hybrid:Korppi2013}.
\begin{figure}
\sidecaption
\includegraphics[width=7.5cm]{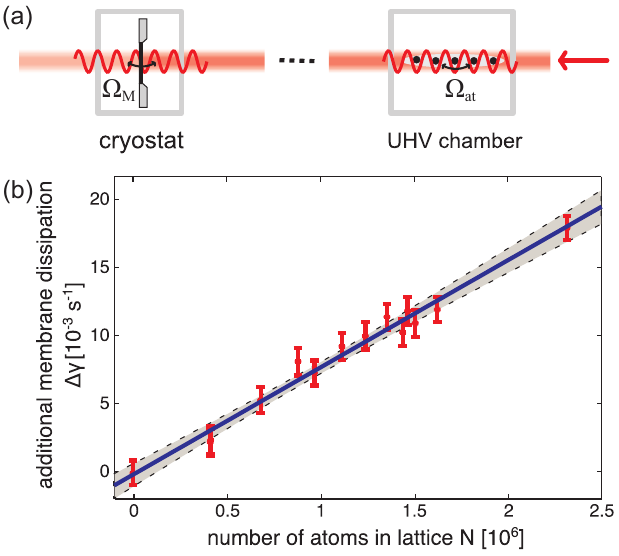}
\caption{(a) Setup: Reflection of light from a micromechanical membrane results in a standing wave which provides an optical lattice potential for ultracold atoms. \index{cold atoms} This configuration gives rise to a coupling of the membrane vibrations and the center of mass motion of the cloud of atoms. Laser cooling of atoms will sympathetically cool along the membrane \cite{Hybrid:Hammerer2010}. (b) Measured change in the mechanical damping rate due to the sympathetic cooling effect in its dependence on atom number in the lattice \cite{Hybrid:Camerer2011}.}
\label{Hybrid:fig:freespace}       
\end{figure}

Consider the setup shown in Fig.~\ref{Hybrid:fig:freespace}a. Laser light is retroreflected from a partially reflective membrane, resulting in a standing wave light field, which in turn provides an optical lattice potential for a cloud of cold atoms. \index{membrane} \index{optical lattice} \index{ultracold atoms}\index{cold atoms}
When the membrane supports a mechanical degree of freedom its position fluctuations will move the optical lattice, and thereby shake along the atoms. Conversely, position fluctuations of atoms in the potential will couple to the membrane's motion: When an atom is displaced from its potential minimum it will experience a restoring force which is due to transfer of photon momentum to the atom. This change in photon momentum is caused by an unbalancing of power between left and right propagating beams, which ultimately changes also the radiation pressure force on the membrane. Thus, also the membrane vibrates along with the atoms. \index{atom-membrane coupling}

A more quantitative, semi-classical consideration along these lines reveals that the (dimensionless) position and momentum fluctuations of the membrane and the atomic COM motion, $\hat q=(\bh+\bh^\dagger)/\sqrt{2}$, $\hat p=i(\bh^\dagger-\bh)/\sqrt{2}$ and $\hat q_{at}=(\ah_{at}+\ah_{at}^\dagger)/\sqrt{2}$, $\hat p_{at}=i(\ah_{at}^\dagger-\ah_{at})/\sqrt{2}$, respectively, \textit{on average} obey the equations of motion
\begin{eqnarray}
\langle \dot{\hat p}\rangle&=&-\OmegaM \langle \hat q\rangle - 2 r \lambda_N\langle \hat q_{at}\rangle, \quad
\langle \dot{\hat q}\rangle=\OmegaM\langle \hat p\rangle,\nonumber\\
\langle \dot{\hat p}_{at}\rangle&=&-\Omega_{at}\langle \hat q_{at}\rangle -2 \lambda_N\langle \hat q\rangle, \quad
\langle \dot{\hat q}_{at}\rangle=\Omega_{at}\langle \hat p_{at}\rangle.\label{Hybrid:Eq:EoMs}
\end{eqnarray}
Here $\Omega_{at}$ denotes the trap frequency for atoms provided by the optical potential, $\lambda_N$ is the coupling strength between the COM motion of atoms and the membrane, and $r$ is the power reflectivity of the membrane. The semiclassical calculation yields a coupling strength of $\lambda_N=(\Omega_{at}/2)\sqrt{Nm_{at}/\mass}$ as in Eq.~(\ref{Hybrid:Eq:directcouplingCOM}), assuming resonance between the two systems $\OmegaM=\Omega_{at}$. For mechanical frequencies on the order of several 100~kHz this condition is routinely met in state of the art optical lattices. As expected from the discussion in section~\ref{Hybrid:sec:direct} the coupling scales with the mass ratio between an atom and the membrane $m_{at}/\mass$, but it is also collectively enhanced by the number of atoms $N$. Therefore even for a mass ratio $m_{at}/\mass\simeq 10^{-14}$, a large but feasible atom number $N=10^8$ will still give rise to an appreciable coupling $\lambda_N$ on the order of kHz for a trap frequency around one MHz.
Moreover, it was recently shown that 
$\lambda_N$ can be further increased by placing the membrane inside an optical cavity \cite{Hybrid:Vogell2013}. Since the atoms are still trapped in the optical lattice forming \textit{outside} the cavity, the long-distance nature of the coupling is maintained. 

Curiously, the semiclassical consideration outlined above predicts a coupling between atoms and membrane that is stronger in one direction than in the other by a factor given by the power reflectivity $r$. This scaling is in fact confirmed and explained by a full quantum treatment of this system. Starting from a complete Hamiltonian description including the motional degress of freedom of the membrane and atoms, as well as the quantized electromagentic field, it is possible to derive a master equation \index{master equation} for the density matrix $\hat \rho$ of the membrane and atomic COM motion \cite{Hybrid:Hammerer2010}. It has the form
\begin{equation}\label{Hybrid:Eq:MEQ}
\dot{\hat \rho}=-i[\hat H_{sys}-2\lambda_N \hat q_{at}\hat q,\hat \rho]+C\hat \rho+L_m\hat \rho+L_{at}\hat \rho.
\end{equation}
and implies the equations of motion (\ref{Hybrid:Eq:EoMs}) for the mean values. The term $C\hat \rho=\textstyle{-i(1-r)\lambda_N}\big([\hat q,\hat q_{at}\hat \rho]-[\hat \rho \hat q_{at},\hat q]\big)$ is responsible for the asymmetric coupling. Its form is well known in the theory of \textit{cascaded quantum systems} \cite{Hybrid:Gardiner2000, Hybrid:Carmichael1993}, and arises here due to the finite reflectivity of the membrane. \index{cascaded quantum system}
The Lindblad terms $L_m\hat \rho$ and $L_{at}\hat \rho$ correspond to momentum diffusion of, respectively, the membrane and the atomic COM motion. They arise due to vacuum fluctuations of the radiation field giving rise to fluctuations of the radiation pressure force on the membrane and the dipole force on atoms. The full quantum treatment of the system correctly reproduces these well known effects, and shows that the corresponding diffusion rates are well below the rates of other relevant decoherence \index{decoherence} processes in this system: For the membrane mode this is thermal heating at a rate $\Gamma_\mathrm{th}$ due to clamping losses or absorption of laser light, which will heat the mechanical mode to thermal occupation $\nth$ in equilibrium. Atoms on the other hand can be \textit{laser cooled} to the motional ground state by well established techniques such as Raman sideband cooling \cite{Hybrid:Camerer2011}. \index{laser cooling}
The corresponding cooling rate $\gamma_{at,cool}$ is widely tunable and can in fact be significantly larger than $\Gamma_\mathrm{th}$. Note that in contrast to the normal optomechanical situation the cooling rate of atoms can be \textit{switched off} giving rise to a regime of coherent coupling between atoms and the membrane.

This opens up the interesting possibility to \textit{sympathetically cool} the membrane motion via laser cooling of atoms in the lattice. \index{sympathetic cooling}
Adding a corresponding heating term for the membrane and a cooling term for atoms to the master equation (\ref{Hybrid:Eq:MEQ}), and solving for the steady state it is possible to determine the effect of the atom-membrane coupling and the associated sympathetic cooling. In \cite{Hybrid:Hammerer2010} it was shown that ground state cooling \index{ground state cooling} might in fact be within reach. In the weak coupling regime $\lambda_N\ll \gamma_{at,cool}$ we expect in analogy to the treatment of optomechanical sideband cooling that laser cooling of atoms results in an increased effective mechanical damping rate $\GammaEff=\GammaM+4r\lambda_N^2/\gamma_{at,cool}$, where $\GammaM=\OmegaM/Q$ is the intrinsic damping rate of the membrane. Such an increase was observed in the experiment reported in \cite{Hybrid:Camerer2011}. Fig.~\ref{Hybrid:fig:freespace}b shows the results for the change in mechanical damping $\Delta\gamma=\GammaEff-\GammaM$ for various experiments with different number of atoms $N$ in the lattice. The linear scaling as expected from the model discussed here is clearly confirmed. Also the magnitude of  $\Delta\gamma$ in the experiment agrees well with the prediction by this model.

Overall this setup provides exciting first results and perspectives for interfacing micromechanical oscillators with ultracold atoms. The interface works at a distance, easing experimental requirements, and enables sympathetic cooling towards the ground state, as well as coherent dynamics for quantum state preparation and measurement of the mechanical mode via coupling to ultracold atoms.

\subsection{Cavity-Optomechanical Coupling Schemes}

The optomechanical coupling discussed in the previous section can be enhanced by placing the atoms and the oscillator inside a high-finesse optical cavity.
In such a system, the cavity field can mediate an interaction between the internal or motional degrees of freedom of the atoms
and the vibrations of the oscillator. In the following, we first discuss a scheme where the oscillator interacts with the internal state of an atomic ensemble. Subsequently, we present a system where the motion of a single trapped atom is strongly coupled to a membrane oscillator. \index{cavity-mediated coupling}\index{optomechanical coupling}


\subsubsection{Coupling of Atomic Internal Levels to Resonator Motion}

The presence of an ensemble of $N$ two-level atoms inside a driven
optical cavity modifies the cavity response. \index{atomic ensemble}
For example when the
atomic resonance is far from the cavity mode frequency, i.e.\ in the
dispersive limit, the effect of the atoms onto the field is a phase
shift. Similarly a vibrating cavity end-mirror also leads to a phase
shift of the intracavity field. The field  can then be used as a
mediator between atoms and mechanical resonator to either allow an
exchange of quantum states, entangle the two systems, or lead to
enhanced optical cooling of the mirror. The last motive will be
explored in the following and it can be seen as an atom induced
effect of spectral filtering of the mirror scattered optical
sidebands. The upshot is that low finesse cavities can provide
resolved sideband cooling when supplemented with filtering ensembles
of atoms \cite{Hybrid:Genes2009}.

The total Hamiltonian of the system can be split as
$\hat H=\hat H_{0}+\hat H_{I}+\hat H_{dis}$, where $\hat H_{0}$ and $\hat H_{I}$ are the free part
and the interaction term, while $\hat H_{dis}$ describes dissipation. The
mirror quadratures are defined as above, $\hat{q}=(\bh+\bh^{\dagger
})/\sqrt 2$ and $\hat{p}=i(\bh^{\dagger}-\bh)/\sqrt 2$, and the
atomic ensemble of frequency splitting $\omega_a$ is described by
creation/annihilation operators $[\hat{c},\hat{c}^{\dagger}]=1$. 
The atom-cavity coupling $G_{a}=g_{a}\sqrt{N}$ is collectively enhanced by $\sqrt{N}$ compared to the single atom-single photon coupling strength $g_{a}$.
The harmonic oscillator description for the atomic cloud is accurate
when the cavity photons are much less numerous than atoms $\ncav\ll
N$. The cavity resonance is $\omegaopt$ and the
laser driving shows up in $\hat H_{I}$ as a displacement term of amplitude $%
\mathcal{E}=\sqrt {2 \Pin \kappa/\hbar \omega_L}$. One can derive
equations of motion from the Hamiltonian dynamics and then perform a
linearization of fluctuations around steady state values that leads
to a set of quantum linearized Langevin equations
\begin{eqnarray}
\dot{\hat{q}}& =&\OmegaM \hat{p}, \\
\dot{\hat{p}}& =&-\GammaM \hat{p}-\OmegaM \hat{q}+\gom(\ah+\ah^{\dagger})+\Langevin, \\
\dot{\ah}& =&-(\kappa +i\Delta _{f})
\ah+i\gom \hat{q}-iG_{a}\hat{c}+\ah_{in}, \\
\dot{\hat{c}}& =&-(\gamma_{a}+i\Delta_{a})\hat{c}-iG_{a}\ah+\hat{c}_{in}.
\end{eqnarray}
where $\Delta_f=\omegaopt-\omega_L-g^2/\OmegaM$ is the effective
cavity detuning, $\kappa$ is the cavity decay rate, $\gamma_{a}$ and $\Delta_{a}= \omega_a-\omega_L$ are the atom
decay rate and detuning respectively, and $\Langevin$, $\hat{a}_{in}$, $\hat{c}_{in}$ is  quantum noise describing the effect of $\hat H_{dis}$ in the Langevin approach.

\begin{figure}[t]
\begin{center}
\includegraphics[width=0.7\columnwidth]{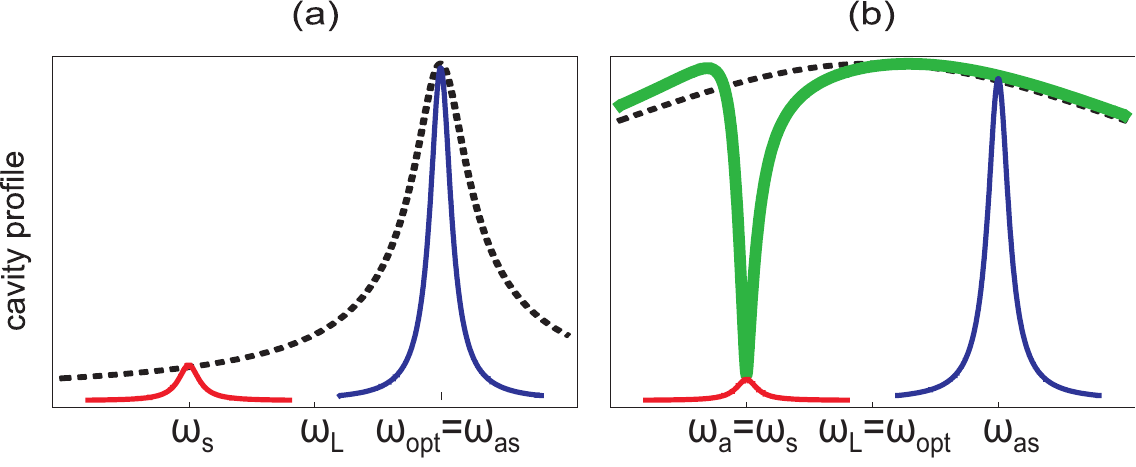}
\end{center}
\caption{Resolved sideband cooling of an end-mirror a) using a sharp
resonance optical cavity and b) using a bad-cavity with an ensemble
of atoms inside. The dotted black lines show the empty cavity
response while the red and blue curves represent optical sidebands.
The green line shows the modification of the cavity response in the
presence of atoms.}
\label{Hybrid:fig:341}       
\end{figure}

Setting $G_{a}=0$ we recover the typical resolved sideband regime in
optomechanics where under the conditions $\kappa \ll \OmegaM$ and $
\Delta _{f}=\OmegaM$, optimal cooling of the mirror via the field is
obtained as illustrated in Fig.~\ref{Hybrid:fig:341}a. The scattered
sidebands are at $\omega _{as,s}=\omega_L\pm \OmegaM$ and the sharp
response of the cavity around $\omegaopt$ leads to a suppression of
the heating sideband. Assuming a bad cavity that cannot resolve
sidebands and $G_{a}>0$, we are instead in the situation depicted in
Fig.~\ref{Hybrid:fig:341}b where the atoms, placed at $\Delta
_{a}=-\OmegaM$, induce a dip in the cavity profile at $\omega _{s}$
that inhibits scattering into the Stokes sideband. Defining atomic
cooperativity $C=G_{a}^{2}/\kappa \gamma _{a} $, the dip at $\omega
_{s}$ scales as $(1+C)^{-1}$. \index{cooperativity}
The width of the dip is $ \gamma
_{a}(1+C)$ representing the enhanced light-induced atomic linewidth.
When the Stokes sideband fits inside the dip one expects an
inhibition by a factor of the order of $(1+C)^{-1}$. For a rigorous
analysis one analyzes the spectrum of the Langevin force
$\hat F=\gom(\ah_{in}+\ah_{in}^{\dagger})$ which gives the cooling and
heating rates (for $\Delta_f=0$): $A_{as}\simeq \gom^{2}/\kappa $
and\ $ A_{s}\simeq \gom^{2}/[\kappa(1+C)] $. Subsequently, the
effective atom-mediated optical damping is
\begin{equation}
\GammaOpt=\frac{\gom^{2}}{\kappa}\frac{C}{1+C}.
\end{equation}

The residual occupancy is given by
$n^{res}=A_{s}/(\GammaM+\GammaOpt)\rightarrow C^{-1}$ (in the limit
of large $C$) and can be compared with $n^{res}=(\kappa
/2\OmegaM)^{2}$ for the purely optomechanical system; an immediate
advantage of this hybrid system comes from the scaling of $n^{res}$
with the controllable parameter $N$.

While in the above the atomic system has been viewed rather as a
high-Q system that acts as a spectral filter for light to relax the
resolved sideband limit requirements, we nevertheless stress that
one can as well take a different stand by seeing the cavity mode
rather as a mediator between atoms and mechanical resonator. From
here, following the elimination of the cavity field as a fast
variable, one can also derive the exact form of the implicit
atom-mirror interaction and show effects such as a quantum state swap
or entanglement \cite{Hybrid:Genes2008}. \index{entanglement}\index{state transfer}

\subsubsection{Coupling of Atomic Motion to Resonator Motion}

\begin{figure}[t]
\sidecaption[t]
\includegraphics[width=7.5cm]{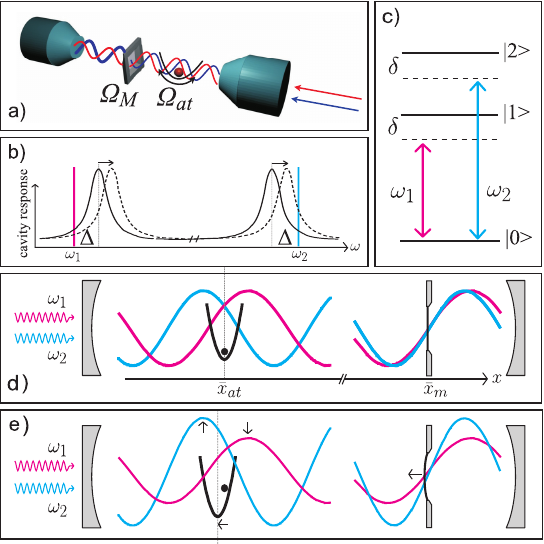}
\caption{Linear atom-membrane coupling mediated by two driven cavity
modes. (a) Schematics of the setup with atom, membrane and two
cavity fields shown. (b) Two cavity resonances and the frequency
position of the two driving lasers. (c)
Internal structure of the atom. (d) Static illustration
of optical potentials. (e) Dynamical illustration of potentials
showing the modification induced by motion of atom and membrane. } \label{Hybrid:fig:343}
\end{figure}

\index{membrane}\index{atom-membrane coupling}
In a situation as that depicted in Fig.~\ref{Hybrid:fig:343}a, where a
single atom is trapped inside a cavity that surrounds a vibrating
membrane, the effect of light on atomic motion can be exploited to
generate motion-motion coupling between membrane and atom
\cite{Hybrid:Hammerer2009b,Hybrid:Wallquist2010}. To this
purpose we choose two  fields of slightly different
wavelengths and opposite detunings $\pm\Delta$ with respect to two cavity
resonances (see Fig.~\ref{Hybrid:fig:343}b) that provide harmonic
trapping of the atom. The two fields have frequencies
$\omega_1,\omega_2$ equally far-detuned by $\delta$ from two
internal transitions of the atom as shown in Fig.~\ref{Hybrid:fig:343}c. Around
the atomic equilibrium position $\bar{x}_{at}$ a Lamb-Dicke
expansion leads to a linear atom-field interaction

\begin{equation}
\hat H_{at,f}=g_{at,f}[(\ah_1+\ah^{\dagger}_1)-(\ah_2+\ah^{\dagger}_2)](\ah_{at}+\ah^{\dagger}_{at}),
\end{equation}
where the atom-field coupling $g_{at,f}$ as well as the atomic 
trapping frequency $\Omega_{at}$ inside the two-field optical trap
are quantities depending on the single-photon Stark shift, cavity
photon number and the geometry of the fields at position
$\bar{x}_{at}$. This interaction can be interpreted as follows (Fig.~\ref{Hybrid:fig:343}d+e):
fluctuations in the amplitudes of the two cavity fields exert
oppositely oriented forces on the atom. Conversely, fluctuations of
the atom around its mean position cause changes of opposite sign in
the amplitudes of the two cavity fields. 
The membrane-field interaction takes a similar form

\begin{equation}
\hat H_{m,f}=g_{m,f}[(\ah_1+\ah^{\dagger}_1)-(\ah_2+\ah^{\dagger}_2)](\bh+\bh^{\dagger}),
\end{equation}
with a similar interpretation. To the reversible Hamiltonian
dynamics one has of course to add the dissipation channels: cavity
decay, momentum diffusion owing to spontaneous scattering of the
atom and thermal decoherence of the membrane. 

The elimination of the fast varying cavity fields can be done in the
limit $|\Delta|\gg g_{at,f},g_{m,f}$, and the reduced atom-membrane
dynamics is governed by a linear two-mode Hamiltonian

\begin{equation}
\hat H_{at,m}=\OmegaM \bh^{\dagger}\bh+\Omega_{at}
\ah^{\dagger}_{at}\ah_{at}-\lambda(\ah_{at}+\ah^{\dagger}_{at})(\bh+\bh^{\dagger}),
\end{equation}
with atom-membrane coupling strength
\begin{equation}
\lambda=\frac{2g_{at,f}g_{m,f}(\Delta+\OmegaM)}{\kappa^2+(\Delta+\OmegaM)^2}+\frac{2g_{at,f}g_{m,f}(\Delta-\OmegaM)}{\kappa^2+(\Delta-\OmegaM)^2},
\end{equation}
to which decoherence at rates \index{decoherence}
$\Gamma_{c}$, $\Gamma_{at}$ and $\Gamma_\mathrm{th}$ adds irreversible dynamics.

The goal is to obtain a coupling $\lambda$ much larger than the
rates of decoherence. For a demonstrative example we consider a
single Cs atom and a SiN membrane of small effective mass
$\mass=0.4$\,ng inside a cavity of finesse of $\finesse\simeq
2\times 10^5$. A small cavity waist of $w_0=10~\mu$m results in a
cooperativity parameter of $140$. With a mechanical quality factor
$Q=10^7$, resonance frequency $\OmegaM=2\pi\times 1.3~$MHz,
circulating power $P_c\simeq 850\,\mu$W and cavity length
$\cavlength=50\,\mu$m we find a cavity mediated coupling
$\lambda\simeq2\pi\times45\,$kHz and decoherence rates
$\Gamma_c,\,\Gamma_\mathrm{th},\,\Gamma_{at}\simeq 0.1\times \lambda$. It it
thus  possible to enter the strong coupling regime with state-of-the-art experimental parameters, even with just a single atom in the cavity. \index{single atom}\index{strong coupling regime}

\section{Conclusion and Outlook}
\label{Hybrid:sec:outlook}

We have discussed various hybrid systems in which a mechanical oscillator is coupled to another (microscopic) quantum system. The approaches that are being pursued are quite diverse, involving superconducting qubits, single spins in the solid state, quantum dots, ultracold atoms in magnetic and optical traps, as well as trapped ions and molecules. One motivation for  building  such hybrid systems is that they enable novel ways to read out and control mechanical objects. For example, a switchable, linear coupling of a mechanical oscillator to a two-level system allows for the preparation of arbitrary quantum states of the oscillator through the Law-Eberly protocol \cite{Hybrid:Law1996}.

Experimentally, the coupling of superconducting two-level systems to mechanical oscillators is most advanced. First experiments have already reached the strong-coupling regime (see \cite{Hybrid:OConnell2010} and chapter [CLELAND]). However, the coherence time of the involved qubits is very short (nano- to microseconds), and it is thus highly desirable to develop and implement strategies for strong coupling of mechanical oscillators to long-lived qubits such as spins in the solid-state or ultracold atoms in a trap. Strong coupling of mechanical oscillators to solid-state spins can build on the impressive achievements of magnetic resonance force microscopy, which has reached single-spin detection sensitivity already some time ago \cite{Hybrid:Rugar2004}. Recently, a novel system was  realized in which the spin of a  nitrogen vacancy center in diamond was used to sense mechanical motion \cite{Hybrid:Arcizet2011,Hybrid:Kolkowitz2012}. 
In another recent experiment, an ensemble of ultracold atoms in an optical lattice was optically coupled to vibrations of a micromechanical membrane \cite{Hybrid:Camerer2011,Hybrid:Korppi2013}, enabling sympathetic cooling of the membrane through laser-cooled atoms. By enhancing the coupling with a high-finesse cavity, strong coupling could be achieved even for a single atom \cite{Hybrid:Hammerer2009b}. \index{strong coupling regime}

The fact that very different microscopic quantum systems are investigated as potential candidates for strong coupling to mechanical oscillators points to one of the big strengths of mechanical quantum systems: the oscillator can be functionalized with electrodes, magnets, or mirrors while maintaining high mechanical quality factor. Mechanical oscillators are thus particularly well suited to serve as quantum transducers \cite{Hybrid:Rabl2010} for precision sensing or hybrid quantum information processing \cite{Hybrid:Wallquist2009}. \index{quantum transducer} \index{hybrid quantum information processing}
Through the mechanical vibrations, spin dynamics can e.g.\ be transduced into electric or optical signals, and one can envision scenarios where atomic quantum memories are interfaced with superconducting quantum processors. 
Another important application is the transduction of microwave or radio-frequency signals into optical signals \cite{Hybrid:Bagci2013,Hybrid:Andrews2013}, ultimately at the level of single quanta \cite{Hybrid:Taylor2011,Hybrid:McGee2013}. 
The versatility of mechanical devices makes them a fascinating toy in the playground of quantum science and technology and we expect many exciting developments in the future.

\acknowledgement{P.T.\ acknowledges fruitful discussions with D. Hunger and support by the EU project AQUTE and the Swiss National Science Foundation. K.H.\ acknowledges support through the cluster of excellence QUEST at the University of Hannover.}

\bibliographystyle{spphys}
\bibliography{Hybrid}

\end{document}